\documentclass[aps,prl,twocolumn,superscriptaddress,showpacs,floatfix]{revtex4}

\usepackage{graphicx}% Include figure files
\usepackage{dcolumn}% Align table columns on decimal point
\usepackage{bm}% bold math

\begin{document}

%\preprint{}

\title{Measurement of the Top-Quark Mass in All-Hadronic Decays \\ in
$p \bar{p}$ Collisions at CDF\,II}
\affiliation{Institute of Physics, Academia Sinica, Taipei, Taiwan 11529, Republic of China} 
\affiliation{Argonne National Laboratory, Argonne, Illinois 60439} 
\affiliation{Institut de Fisica d'Altes Energies, Universitat Autonoma de Barcelona, E-08193, Bellaterra (Barcelona), Spain} 
\affiliation{Baylor University, Waco, Texas  76798} 
\affiliation{Istituto Nazionale di Fisica Nucleare, University of Bologna, I-40127 Bologna, Italy} 
\affiliation{Brandeis University, Waltham, Massachusetts 02254} 
\affiliation{University of California, Davis, Davis, California  95616} 
\affiliation{University of California, Los Angeles, Los Angeles, California  90024} 
\affiliation{University of California, San Diego, La Jolla, California  92093} 
\affiliation{University of California, Santa Barbara, Santa Barbara, California 93106} 
\affiliation{Instituto de Fisica de Cantabria, CSIC-University of Cantabria, 39005 Santander, Spain} 
\affiliation{Carnegie Mellon University, Pittsburgh, PA  15213} 
\affiliation{Enrico Fermi Institute, University of Chicago, Chicago, Illinois 60637} 
\affiliation{Comenius University, 842 48 Bratislava, Slovakia; Institute of Experimental Physics, 040 01 Kosice, Slovakia} 
\affiliation{Joint Institute for Nuclear Research, RU-141980 Dubna, Russia} 
\affiliation{Duke University, Durham, North Carolina  27708} 
\affiliation{Fermi National Accelerator Laboratory, Batavia, Illinois 60510} 
\affiliation{University of Florida, Gainesville, Florida  32611} 
\affiliation{Laboratori Nazionali di Frascati, Istituto Nazionale di Fisica Nucleare, I-00044 Frascati, Italy} 
\affiliation{University of Geneva, CH-1211 Geneva 4, Switzerland} 
\affiliation{Glasgow University, Glasgow G12 8QQ, United Kingdom} 
\affiliation{Harvard University, Cambridge, Massachusetts 02138} 
\affiliation{Division of High Energy Physics, Department of Physics, University of Helsinki and Helsinki Institute of Physics, FIN-00014, Helsinki, Finland} 
\affiliation{University of Illinois, Urbana, Illinois 61801} 
\affiliation{The Johns Hopkins University, Baltimore, Maryland 21218} 
\affiliation{Institut f\"{u}r Experimentelle Kernphysik, Universit\"{a}t Karlsruhe, 76128 Karlsruhe, Germany} 
\affiliation{High Energy Accelerator Research Organization (KEK), Tsukuba, Ibaraki 305, Japan} 
\affiliation{Center for High Energy Physics: Kyungpook National University, Taegu 702-701, Korea; Seoul National University, Seoul 151-742, Korea; and SungKyunKwan University, Suwon 440-746, Korea} 
\affiliation{Ernest Orlando Lawrence Berkeley National Laboratory, Berkeley, California 94720} 
\affiliation{University of Liverpool, Liverpool L69 7ZE, United Kingdom} 
\affiliation{University College London, London WC1E 6BT, United Kingdom} 
\affiliation{Centro de Investigaciones Energeticas Medioambientales y Tecnologicas, E-28040 Madrid, Spain} 
\affiliation{Massachusetts Institute of Technology, Cambridge, Massachusetts  02139} 
\affiliation{Institute of Particle Physics: McGill University, Montr\'{e}al, Canada H3A~2T8; and University of Toronto, Toronto, Canada M5S~1A7} 
\affiliation{University of Michigan, Ann Arbor, Michigan 48109} 
\affiliation{Michigan State University, East Lansing, Michigan  48824} 
\affiliation{Institution for Theoretical and Experimental Physics, ITEP, Moscow 117259, Russia} 
\affiliation{University of New Mexico, Albuquerque, New Mexico 87131} 
\affiliation{Northwestern University, Evanston, Illinois  60208} 
\affiliation{The Ohio State University, Columbus, Ohio  43210} 
\affiliation{Okayama University, Okayama 700-8530, Japan} 
\affiliation{Osaka City University, Osaka 588, Japan} 
\affiliation{University of Oxford, Oxford OX1 3RH, United Kingdom} 
\affiliation{University of Padova, Istituto Nazionale di Fisica Nucleare, Sezione di Padova-Trento, I-35131 Padova, Italy} 
\affiliation{LPNHE, Universite Pierre et Marie Curie/IN2P3-CNRS, UMR7585, Paris, F-75252 France} 
\affiliation{University of Pennsylvania, Philadelphia, Pennsylvania 19104} 
\affiliation{Istituto Nazionale di Fisica Nucleare Pisa, Universities of Pisa, Siena and Scuola Normale Superiore, I-56127 Pisa, Italy} 
\affiliation{University of Pittsburgh, Pittsburgh, Pennsylvania 15260} 
\affiliation{Purdue University, West Lafayette, Indiana 47907} 
\affiliation{University of Rochester, Rochester, New York 14627} 
\affiliation{The Rockefeller University, New York, New York 10021} 
\affiliation{Istituto Nazionale di Fisica Nucleare, Sezione di Roma 1, University of Rome ``La Sapienza," I-00185 Roma, Italy} 
\affiliation{Rutgers University, Piscataway, New Jersey 08855} 
\affiliation{Texas A\&M University, College Station, Texas 77843} 
\affiliation{Istituto Nazionale di Fisica Nucleare, University of Trieste/\ Udine, Italy} 
\affiliation{University of Tsukuba, Tsukuba, Ibaraki 305, Japan} 
\affiliation{Tufts University, Medford, Massachusetts 02155} 
\affiliation{Waseda University, Tokyo 169, Japan} 
\affiliation{Wayne State University, Detroit, Michigan  48201} 
\affiliation{University of Wisconsin, Madison, Wisconsin 53706} 
\affiliation{Yale University, New Haven, Connecticut 06520} 
\author{T.~Aaltonen}
\affiliation{Division of High Energy Physics, Department of Physics, University of Helsinki and Helsinki Institute of Physics, FIN-00014, Helsinki, Finland}
\author{A.~Abulencia}
\affiliation{University of Illinois, Urbana, Illinois 61801}
\author{J.~Adelman}
\affiliation{Enrico Fermi Institute, University of Chicago, Chicago, Illinois 60637}
\author{T.~Affolder}
\affiliation{University of California, Santa Barbara, Santa Barbara, California 93106}
\author{T.~Akimoto}
\affiliation{University of Tsukuba, Tsukuba, Ibaraki 305, Japan}
\author{M.G.~Albrow}
\affiliation{Fermi National Accelerator Laboratory, Batavia, Illinois 60510}
\author{D.~Ambrose}
\affiliation{Fermi National Accelerator Laboratory, Batavia, Illinois 60510}
\author{S.~Amerio}
\affiliation{University of Padova, Istituto Nazionale di Fisica Nucleare, Sezione di Padova-Trento, I-35131 Padova, Italy}
\author{D.~Amidei}
\affiliation{University of Michigan, Ann Arbor, Michigan 48109}
\author{A.~Anastassov}
\affiliation{Rutgers University, Piscataway, New Jersey 08855}
% ...
\collaboration{CDF Collaboration\footnote{With visitors from $^a$University of Athens, 
$^b$University of Bristol, 
$^c$University Libre de Bruxelles, 
$^d$Cornell University, 
$^e$University of Cyprus, 
$^f$University of Dublin, 
$^g$University of Edinburgh, 
$^h$University of Heidelberg, 
$^i$Universidad Iberoamericana, 
$^j$University of Manchester, 
$^k$Nagasaki Institute of Applied Science, 
$^l$University de Oviedo, 
$^m$University of London, Queen Mary and Westfield College, 
$^n$Texas Tech University, 
$^o$IFIC(CSIC-Universitat de Valencia), 
}}
\noaffiliation


%\author{}
%\thanks{}
%\affiliation{}


\date{\today}

\begin{abstract}
We present a measurement of the top-quark mass, $M_{\mathrm{top}}$, in the
all-hadronic decay channel $t\bar{t}
\to W^+b W^- \bar{b} \to q_1\bar{q}_2 b  q_3 \bar{q}_4 \bar{b}$.
The analysis is performed using 310\,pb$^{-1}$ of $\sqrt{s}$=1.96 TeV
$p\bar{p}$ collisions collected with the CDF\,II detector using a
multi-jet trigger.
The mass measurement is based on an event-by-event likelihood which depends on both the
sample purity and the value of the top-quark mass, using 90 possible
jet-to-parton assignments in the six-jet final state.
The joint likelihood of 290 selected events yields a value of
$M_{\mathrm{top}}$=177.1 $\pm$ 4.9 (stat.) $\pm$ 4.7 (syst.)~GeV/$c^2$. 
\end{abstract}

\pacs{14.65.Ha, 12.15.Ff}
%\keywords{CDF/PHYS/TOP/PUBLIC/8491 \\ PRL Draft 3}

\maketitle

\clearpage
\newpage
The mass of the top quark, $M_{\mathrm{top}}$,  is an important free parameter of the standard model  (SM) and is of the order of the electroweak 
symmetry breaking scale. Since virtual top quarks are involved in
higher-order electroweak processes, by measuring the top-quark mass one
can constrain the mass of the SM Higgs
boson~\cite{ewgroup} and particles predicted in extensions of the SM~\cite{susy}.
At the Tevatron, $t\bar{t}$  pairs are produced by QCD processes and
decay according to the CKM matrix~\cite{pdg} with a branching ratio of almost 100\%  into
a $W$ boson and a $b$ quark.
The final state of the event is then defined by the decay of the $W$ bosons.
All-hadronic $t\bar{t}$ events, where both of the $W$ bosons
decay into quarks, have a multi-jet final state and no missing energy due to neutrinos.
The top-quark mass measurement in this decay channel 
is motivated by the large branching fraction
($\approx$44\%) and the complete reconstruction of both top and antitop
quarks, relying only on hadronic jets. It is the first top-quark
mass measurement in this channel using Tevatron Run\,II data, and serves as a consistency check of the measurements in
the two other
$t\bar{t}$ decay modes~\cite{dilepton,lplusjet}.
The major experimental challenge is the
presence of a large amount of multi-jet background events from QCD processes, which dominates
$t\bar{t}$ production by three orders of magnitude, even after applying a trigger dedicated to events
with a multi-jet final state.

In this Letter, we present the measurement of the top-quark mass in
the all-hadronic decay channel using a sample of $t\bar{t}$ decays
corresponding to 310 pb$^{-1}$ of proton-antiproton collisions at
$\sqrt{s}$=1.96 TeV, collected using the CDF\,II detector between February, 2002, and August, 2004. 
We measure the mass using the ideogram method, which was
used in the DELPHI experiment at LEP for the $W$ boson mass
measurement~\cite{delphi}. 
The method is based on an event-by-event likelihood that reconstructs the top
and antitop quarks using all 90  assignments
of jets to quarks from $W$ boson decays and $b$ quarks (jet combinations) from the $t\bar{t}$
decay. Considering all jet combinations enhances the amount of top-quark mass
information we can extract from each event.
 The 90 jet combinations arise from the 6!/8 permutations of the six
jets, where the two possible top antitop permutations and the two possible
permutations of each of the two quark pairs coming from the $W$ decay are
treated equally.

 The sample purity, $\mathcal{P}$, defined as the fraction of all-hadronic $t\bar{t}$
events contained in the selected data sample, is a free parameter of the likelihood
together with the value of $M_{\mathrm{top}}$. The reason for measuring both
$M_{\mathrm{top}}$ and $\mathcal{P}$ simultaneously is that the cross
sections of the QCD backgrounds are not well known. Even though the background
processes do not contain any the top-quark mass information, the
sensitivity of the likelihood on $M_{\mathrm{top}}$ does depend on the
total amount of QCD background events. This is reflected
both in the statistical variance of $M_{\mathrm{top}}$, and in corrections
applied to $M_{\mathrm{top}}$ which take into account the presence of background events.
%The basic idea is to abandon the analysis based on fitting the 
%line shape to the global invariant mass spectrum, and to use instead
%event-by-event likelihoods describing the full ambiguity of the mass
%information in each event as correctly as possible. 

The CDF\,II detector~\cite{CDFdet} is a general-purpose charged and neutral
particle detector designed to study $p\bar{p}$ collisions at the Fermilab
Tevatron. It consists of an eight-layer silicon microstrip detector array and
 a cylindrical drift chamber contained inside a {\sloppy 1.4\,-T} solenoid magnet, 
surrounded by  electromagnetic and hadronic sampling
calorimeters with a geometrical acceptance up to a pseudorapidity 
of $|\eta|$ = 3.6~\cite{frame}. Muon chambers are located outside the calorimeters,
allowing the reconstruction of track segments for penetrating particles. 

The dataset used for this measurement is selected with a  multi-jet
trigger that relies solely on calorimetry. This trigger requires at least
four calorimeter clusters with a transverse energy $E_\mathrm{T} > 15$~GeV
and the scalar sum of the transverse jet energies to exceed 125~GeV. 
To improve the signal-to-background
ratio, we impose offline kinematic
requirements based on the scalar sum of the transverse jet energies 
and event shape observables.  
A full description of the trigger and kinematic selection is
given in an earlier publication~\cite{allhad}. 
For the final top-quark mass measurement we consider only events with exactly six jets,
each with
transverse energy $E_{\mathrm{T}} > 15$ ~GeV and pseudorapidity $|\eta| \leq 2$.
Jets are identified as clusters
of energy deposits in the calorimeter segments (towers)  which fall within a cone radius of $\Delta R = \sqrt{\Delta \phi^2 +
\Delta \eta^2} \leq 0.4$~\cite{frame}. 
The raw jet energy scale (JES) is the result of a multi-step
correction procedure to convert the measured transverse jet energy to the
expected transverse energy of the parton corresponding to the jet. 
The corrections, assessed using data and a simulation of the CDF detector, 
include corrections for the response inhomogeneity in $\eta$,
contributions from multiple interactions, the non-linearity of the
calorimeter response, the underlying event, and the energy flow out of the jet cone~\cite{jets}.
%A correction to the jet energy scale (JES),
%assessed using a simulation of the CDF detector, is employed to
%convert the measured transverse jet energy to the expected
%transverse energy of the parton corresponding to the jet. 
%This is complemented by additional
%corrections for the calorimeter energy non-linearity and the response inhomogeneity in $\eta$, multiple
%interactions, the underlying event, and the energy flow out of the jet
%cone~\cite{jets}. 
We apply additional jet-parton energy corrections, specific to $t\bar{t}$ events, 
which are parameterized independently for $b$-quark jets and
light-flavor jets~\cite{flavor}. 
%(so called top specific corrections).

In order to further reduce the
QCD background, we apply $b$-quark identification to each event passing the kinematic
selection. At least one jet must have a reconstructed displaced vertex ($b$-tag), consistent with
a long-lived bottom hadron~\cite{btag}. This displaced vertex information is not used to reduce the number of jet combinations in the event
likelihood. Instead, we derive a weight factor using the CDF jet probability
algorithm~\cite{jetprob} which takes into account the probability that
a jet originates from a $b$ quark, as explained below.
The final data sample contains 290 events. Assuming a theoretical cross section of 6.1 pb~\cite{crosssec} 
for a top-quark mass of 178~GeV/$c^2$~\cite{run1comb}, the expected signal-to-background
ratio is about one-to-five. 

By applying energy and momentum conservation we fit the momenta of the jets and the two top-quark masses,
$m_i^{1,2}$, and determine their estimated uncertainties,
$\sigma_i^{1,2}$ using a $\chi^2$ minimization. This kinematic fit is applied
to each of the 90 jet combinations, $i$, where we constrain the  masses
of the two $W$ bosons to a world average value of
$m_W=80.4$~GeV/$c^2$~\cite{pdg} within their natural widths. 
In order to achieve a better background
reduction, the two reconstructed top-quark masses,
$m_i^{1,2}$, are not constrained to be equal. The $m_i^{1,2}$
distributions thus obtained peak around 160~GeV/$c^2$ and range from 100--300~GeV/$c^2$ for the whole data sample, while the
uncertainties $\sigma_i^{1,2}$ are typically of the order of 15~GeV/$c^2$ and
range from 5--30~GeV/$c^2$. 
We measure $M_{\mathrm{top}}$ using an event likelihood where we weigh each combination $i$ 
with a factor $w_i$, expressing the compatibility with both the $t\bar{t}$
decay kinematics and the presence of two $b$-quark jets together with four
light-flavor jets. The weight factor $w_i$ is calculated as
\begin{equation}
w_i=\exp{(-\frac{1}{2}\chi^2_i)} w_i^{b},
\end{equation}
where the exponential term, calculated using the $\chi^2$ value
of the kinematic fit, is a measure of the compatibility of jet
combination $i$ with the kinematics of a decaying $t\bar{t}$ pair. The second
factor, $w_i^b$, gives the probability of two of the jets to originate from a $b$
(or $\bar{b}$) quark and four others to originate from light quarks
(including $c$ quarks), and is given by
\begin{equation}
w_i^{b}=\prod_{j=1}^{2}p^b_j \prod_{j=3}^{6}p^q_j,
\label{jpweight}
\end{equation}
where we assume quarks with indices 3-6  to be light quarks originating from
the $W$ bosons.
The probability $p_j^{b({\rm or}~q)}$ for a jet $j$ to originate from a
$b$ quark (or light-flavor quark, $q$) is obtained using the CDF jet probability
algorithm~\cite{jetprob}. This calculates the
probability $P_j$ for a given jet $j$ to originate from the primary vertex, based
on the impact parameter information from the tracks belonging to that jet.
The $P_j$ distribution for
$b$-quark jets (or light-flavor jets) is described by a probability density
function, $f(b({\rm or}~q)|P_j)$ in order to obtain
\begin{equation}
p^{b({\rm or}~q)}_j=\frac{f(b({\rm or}~q)|P_j)}{f(b|P_j)+f(q|P_j)}.
\end{equation}

The event likelihood
consists of two terms.
The signal likelihood term corresponds to the convolution
of two Breit-Wigner distributions, $F_\mathrm{BW}(m_j'|M_{\mathrm{top}})$, with
two Gaussians, $G(m_j'|m_i^{j},\sigma_i^{j})$, describing the experimental
resolutions, $\sigma_i^{1,2}$, for each of the two reconstructed top-quark masses, $m_i^{1,2}$:
\begin{eqnarray}
\mathcal{L}_i^{\mathrm{sig}}(M_{\mathrm{top}})=&&\prod_{j=1,2}\int
G(m_j'|m_i^{j},\sigma_i^{j}) \nonumber \\&& \times F_\mathrm{BW}(m_j'|M_{\mathrm{top}})\mathrm{d}m_j'.
\end{eqnarray}
The $M_{\mathrm{top}}$-independent background likelihood term,
$\mathcal{L}_i^{\mathrm{bg}}$, corresponds to the two-dimensional posteriori probability
density function (histogram) of $m_i^{1,2}$ obtained from {\sc alpgen}~\cite{alpgen}
Monte Carlo (MC) multi-jet QCD  background. All MC events
are passed through the CDF detector simulation and are subjected to the same
event selection criteria as the data.

The likelihood for a given event $n$ is derived by summing the signal and background event
likelihoods for each jet combination $i$ and is calculated as a function of the top-quark
mass, $M_{\mathrm{top}}$, and the sample purity, $\mathcal{P}$:
\begin{equation}
\mathcal{L}^{n}(M_{\mathrm{top}},\mathcal{P})=\sum_{i=1}^{90}w_i\bigl[\mathcal{P}
\mathcal{L}_i^{\mathrm{sig}}(M_{\mathrm{top}})+(1-\mathcal{P}) \mathcal{L}_i^{\mathrm{bg}}\bigr]. \label{eq1}
\end{equation}

%As an illustration of the top-quark mass spectrum, Fig.~\ref{proplot} shows
%the distribution of the top-quark mass values 
%corresponding to the maximum of the likelihood of each event 
%for data and MC expectations. The $t\bar{t}$
%signal events are generated with {\sc herwig}~\cite{herwig}.
%, using a
%generator top-mass value of 178.0~GeV/$c^2$.
%The event generator {\sc HERWIG}~\cite{herwig} is used to generate $t\bar{t}$
%signal event samples over a top-quark mass range between 160~GeV/$c^2$ and 190
%GeV/$c^2$. 
%The QCD background is modeled using {\sc alpgen} MC
%events. The false identification of $b$-quarks is modeled
%using the tag rate parameterization used in~\cite{allhad}. Note that Fig.~\ref{proplot}  is a
%demonstration of the agreement between data and MC expectation and is not used in the
%final measurement of the top quark mass, since it does not contain the full
%information contained in the actual likelihood fit. 
We obtain a one-dimensional likelihood curve as a function
 of $M_{\mathrm{top}}$ by maximizing the two-dimensional joint likelihood, 
\begin{equation}
L(M_{\mathrm{top}},\mathcal{P})=\prod_{n=1}^{290}
\mathcal{L}^{n}(M_{\mathrm{top}},\mathcal{P}),
\label{jl}
\end{equation}
with respect to the sample
 purity, $\mathcal{P}$, for each value of $M_{\mathrm{top}}$. 
%This yields a one-dimensional likelihood curve as function
% of $M_{\mathrm{top}}$. 
The maximum of the total likelihood of the 290 selected data events
 corresponds to a sample purity value of $\mathcal{P}=0.21\pm 0.07
 \rm{(stat.)}$, which is compatible with a signal-to-background ratio of about one-to-five expected from SM $t\bar{t}$
 production.
%Since this figure is not used to finally measure the top-quark mass, some care needs to be
%taken for its interpretation. Figure~\ref{proplot} does not take the width,
%nor the absolute scale of the event likelihoods into account. In addition,
%the event likelihoods are non-Gaussian. As a result, the background MC expectation peaks close to the signal MC
%expectation in Fig.~\ref{proplot} but the maximum of the total background
%likelihood curve peaks at $M_{\mathrm{top}}\approx 182$~GeV/$c^2$ and its maximum is several orders
%of magnitude smaller than the maximum of the signal likelihood.
\begin{figure}[!tbh]
\begin{center}
\includegraphics[width=7.8cm,clip=]{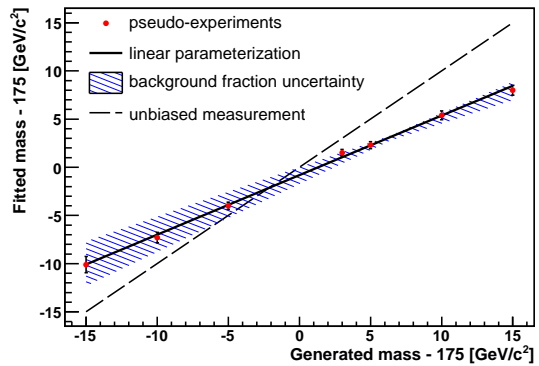}\\
\caption{The correlation between the measured value of $M_{\mathrm{top}}$ and the
generated top-quark mass obtained with pseudo-experiments using $t\bar{t}$
signal events generated with {\sc herwig} and an admixture of QCD background
generated with {\sc alpgen}. The solid line represents the linear
parameterization used to correct $M_{\mathrm{top}}$.
The shaded band shows the effect of a one-sigma
variation of the sample purity, $\mathcal{P}=0.21\pm 0.07
\rm{(stat.)}$. The dashed line corresponds to 100\% correlation and zero offset.\label{calib}}
\end{center}
\end{figure} 
%The signal likelihood term in Eq.~(\ref{eq1}) assumes that the kinematic fit
%returns on average the correct 
The value of $M_{\mathrm{top}}$ extracted from the joint likelihood fit in
Eq.~(\ref{jl}) is biased due to the presence of wrong jet combinations,
background events, and assignment of jets that arise from
initial- (ISR) and final-state gluon radiation (FSR).
Therefore, we correct the measurement of $M_{\mathrm{top}}$ using
a linear parameterization of the correlation between the generated
top-quark mass value and the mass estimator $M_{\mathrm{top}}$ in the range between 160~GeV/$c^2$ and 190
GeV/$c^2$, as shown in Fig.~\ref{calib}. Each mass point is obtained from pseudo-experiments with MC samples containing a
mixture of 290 Poisson-fluctuated signal plus background events. 
We use {\sc herwig}~\cite{herwig} to generate the $t\bar{t}$ signal
events and {\sc alpgen} to model the QCD background.
The tag rate matrix used in~\cite{allhad} models the false identification of $b$-quarks.
The correction depends on the sample purity used to
construct the pseudo-experiments as illustrated by the shaded band
which corresponds to the one-sigma statistical uncertainty on the sample purity
measured from data. Accordingly, we assign a systematic uncertainty due to
this purity uncertainty.
Using the same pseudo-experiments we verify that the measured sample
purity is independent of the generator top-quark mass value and that the statistical uncertainty on
$M_{\mathrm{top}}$ covers 68\% of the sample measurements. We correct the
statistical uncertainty from the likelihood curve using the
width of the pull distribution
$(M_{\mathrm{top}}-m_{\mathrm{gen}})/\sigma_{M_{\mathrm{top}}}$, where
$m_{\mathrm{gen}}$ is the generated top-quark mass, $M_{\mathrm{top}}$ the
measured top-quark mass, and $\sigma_{M_{\mathrm{top}}}$ its statistical
uncertainty obtained in the pseudo-experiments. The correction factor, 1.17,
is independent of the value of $M_{\mathrm{top}}$.

\begin{figure}[!tbp]
\begin{center}
\includegraphics[width=7.8cm,clip=]{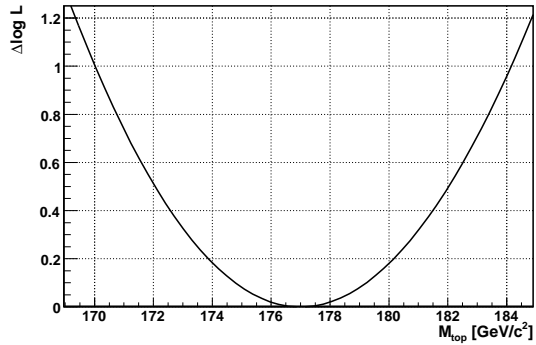}
\caption{The one-dimensional negative log-likelihood of $M_{\mathrm{top}}$
obtained including
corrections using the parameterization shown in Fig.~\ref{calib}\label{likelihood}}
\end{center}
\end{figure}

%Besides the sample purity, the relative fractions of light-flavor QCD background and 
%QCD background containing $b$ quarks could influence the value of
%$M_{\mathrm{top}}$. Their ratio is estimated to be 2.7 by using the ratio of 
%their respective cross sections given by the
%{\sc ALPGEN} event generator and their respective selection efficiencies. 
%This number is verified by a comparison between data and MC
%expectation in the signal region using the jet probability distribution, which is sensitive to the QCD background
%composition. A binned maximum likelihood fit of the ratio
%of the light/$b$-quark flavor QCD background yields  a value of 2.8 $\pm$
%1.0. Although the measurement of $M_{\mathrm{top}}$ is nearly insensitive to the flavor composition of the
%background, a small systematic uncertainty is assigned and described below. 
%%% Add some text
  The maximum of the one-dimensional top mass likelihood, after
 correction with the parameterization shown in Fig.~\ref{calib},
 corresponds to the value  $M_{\mathrm{top}}=177.1 \pm 4.9
 \rm{(stat.)}$~GeV/$c^2$. Figure.~\ref{likelihood} shows the corresponding likelihood.

The uncertainty on the JES which varies
between approximately 3\% and  8\% of the measured jet energy, depending on the $\eta$ and
$p_\mathrm{T}$ of the jet~\cite{jets}, dominates the systematic uncertainties on our measurement of
$M_{\mathrm{top}}$. We determine its effect using pseudo-experiments in which the JES is increased (decreased)
with one standard deviation. Half of the difference in $M_{\mathrm{top}}$
between the positive and negative variation of the JES  amounts to $\Delta
M_{\mathrm{top}} = 4.3$~GeV/$c^2$. 
%However, the uncertainty on the JES is
%correlated with the shape and number of background events.
%We take this into account by reducing the error assigned to the
%JES uncertainty from 4.6~GeV/$c^2$ to 4.3~GeV/$c^2$.

The corrections to $M_{\mathrm{top}}$, shown in Fig.~\ref{calib}, depend both on the background fraction and shape.
The uncertainty due to the background fraction, varied within the uncertainties of the measured sample purity, is $\Delta
M_{\mathrm{top}} = 1.1$~GeV/$c^2$ for the measured value of $M_{\mathrm{top}}$. 
The relative fractions of the light-flavor QCD background and 
QCD background containing $b$ quarks is not precisely known. 
We estimate their ratio (light-flavor/$b$) to be  2.8 $\pm$ 1.0 by
performing a binned maximum likelihood fit of the MC expectation to the data
based on the jet probability weight distribution in Eq.~(\ref{jpweight}).
This result is consistent with a value of 2.7 obtained using the respective
QCD cross sections given by the {\sc alpgen} event generator, taking the
individual selection efficiencies into account. 
A systematic uncertainty due to the background composition results in $\Delta
M_{\mathrm{top}} = 0.8$~GeV/$c^2$ by varying the background
flavor composition between only light-flavor QCD background events and only
QCD background containing $b$ quarks. 

In accordance with~\cite{lplusjet}, we vary two {\sc pythia}~\cite{pythia} parton shower
parameters,  $\Lambda_{\mathrm{QCD}}$ and the ISR/FSR transverse momentum
scale $K$-factor, in order to model the systematic uncertainties due to the amount of ISR and
FSR. 
%The parameters variation ranges are 73--292 MeV for
%$\Lambda_{\mathrm{QCD}}$ and 0.5--2.0 for the $K$-factor.
The largest difference observed in $M_{\mathrm{top}}$ between any 
variation amounts to 0.9~GeV/$c^2$ (0.8
GeV/$c^2$) for ISR (FSR).

To examine the systematic effect due to uncertainties in the parton
distribution function (PDF) of the proton and antiproton, we follow the approach used
in~\cite{lplusjet}, resulting in $\Delta M_{\mathrm{top}} = 0.8$~GeV/$c^2$.
%where 20 pairs of
%$\pm 1\sigma$ uncertainty sets based on CTEQ6M are used~\cite{pdf1,pdf2}. 
%The induced uncertainties are added in quadrature, together with the
%difference in $M_{\mathrm{top}}$ when switching between the leading order PDFs CTEQ5L and MRST72,
%and variations induced by changing $\Lambda_{\mathrm{QCD}}$ in the MRST
%set~\cite{pdf3}. This results in a total PDF uncertainty of 0.82~GeV/$c^2$.

By applying the parameterized corrections on $M_{\mathrm{top}}$
obtained with  {\sc herwig} to MC events
generated with {\sc pythia} we estimate MC modeling uncertainties.  
%In addition to differences in the fragmentation
%model, {\sc HERWIG} and {\sc PYTHIA} treat polarization and helicity effects differently.
The resulting difference in $M_{\mathrm{top}}$ equals 0.5~GeV/$c^2$.

Uncertainties in $b$-quark fragmentation, semi-leptonic branching ratios, and
color-flow affect the $b$-quark JES~\cite{lplusjet}.
This results in an additional contribution to $\Delta M_{\mathrm{top}}$ of  0.5~GeV/$c^2$.
All systematic uncertainties are summarized in Table~\ref{syst}. 
They result in a total systematic uncertainty of $\Delta M_{\mathrm{top}}=4.7$~GeV/$c^2$.
\begin{table}[!t]
\begin{center}
\caption{Systematic uncertainties on the measured top-quark mass.}
\begin{tabular}{lc} \hline\hline
 Systematic \\ uncertainty source& $\Delta M_{\mathrm{top}}$ [GeV/$c^2$] \\
\hline
JES& 4.3\\
Background fraction& 1.1\\
Background shape& 0.8\\
ISR& 0.9\\
FSR& 0.8\\
PDF& 0.8\\
MC modeling (generator) & 0.5\\
$b$-quark JES & 0.5\\\hline
Total & 4.7\\\hline\hline
\end{tabular}
\label{syst}
\end{center}
\end{table}

In summary, we present a new measurement of the top-quark mass in 
the all-hadronic channel. The measured value is $M_{\mathrm{top}}=177.1
\pm 4.9\rm{(stat.)} \pm 4.7\rm{(syst.)}$~GeV/$c^2$. 
This measurement is the first determination of the top-quark mass in the
all-hadronic channel using Run\,II data and is twice as precise as the 
Run\,I measurements in this channel~\cite{run1mass,d0run1}.
We expect the statistical uncertainty of this measurement 
to be reduced to 2~GeV/$c^2$ after analyzing 1\,fb$^{-1}$ of integrated luminosity.
%The clear challenge will obviously be the reduction of the jet
%energy scale uncertainty in future all-hadronic measurements. This can be
%accomplished by e.g. measuring the
%JES simultaneously with $M_{\mathrm{top}}$ using the mass of the hadronically
%decaying $W$ as a constraint.
Our result can be compared with the latest CDF top-quark mass measurements
in the dilepton~\cite{dilepton} and lepton+jets~\cite{lplusjet} channel,
which are based on samples with a similar 
integrated luminosity. The measured value for the dilepton channel is
$M_{\mathrm{top}}=167.9  \pm 5.2 \rm{(stat.)}  \pm  3.7 \rm{(syst.)}$~GeV/$c^2$, 
and the value for the lepton+jets channel is 
$M_{\mathrm{top}}=173.5^{+3.7}_{-3.6} \rm{(stat.+JES)}  \pm  1.3
\rm{(syst.)}$~GeV/$c^2$, where the first uncertainty includes the uncertainty on
the jet energy scale, which is measured simultaneously with  $M_{\mathrm{top}}$.
At this level of precision, the values of $M_{\mathrm{top}}$ are compatible
between all three decay channels. 
%However, the tendency to observe larger
%$M_{\mathrm{top}}$ values  in both the
%lepton+jets and the all-hadronic channels with respect to the dilepton
%measurements may indicate a missing or underestimated source of systematic
%uncertainty which will reveal itself as the precision of the
%top-quark mass measurement in each channel continues to improve. 
%
%estimates of the total
%systematic uncertainties might become inadequate with increasing precision of the
%top-quark mass measurements. 
%Color flow, parton showering effects and
%underlying event activity might be additional sources of systematic
%uncertainties that deserve more careful study in the near future.
%\end{introduction} 

%\begin{acknowledgments}
%check for the latest@
%http://www-cdf.fnal.gov/internal/people/links/CarolPicciolo/standardack.html
%\section{Acknowledgments}
We thank the Fermilab staff and the technical staffs of the participating
institutions for their vital contributions. This work was supported by the
U.S. Department of Energy and National Science Foundation; the Italian
Istituto Nazionale di Fisica Nucleare; the Ministry of Education, Culture,
Sports, Science and Technology of Japan; the Natural Sciences and Engineering
Research Council of Canada; the National Science Council of the Republic of
China; the Swiss National Science Foundation; the A.P. Sloan Foundation; the
Bundesministerium f\"ur Bildung und Forschung, Germany; the Korean Science
and Engineering Foundation and the Korean Research Foundation; the Particle
Physics and Astronomy Research Council and the Royal Society, UK; the
Institut National de Physique Nucleaire et Physique des Particules/CNRS; the
Russian Foundation for Basic Research; the Comisi\'on Interministerial de
Ciencia y Tecnolog\'{\i}a, Spain; the European Community's Human Potential
Programme under contract HPRN-CT-2002-00292; and the Academy of Finland.


\begin{thebibliography}{99}
\bibitem{frame}
The CDF reference frame uses cylindrical coordinates, where $\theta$ and
$\phi$ are the polar and azimuthal angles with respect to the 
proton beam. The pseudorapidity is defined as $\eta=-\ln{\tan{\theta /2}}$,
the transverse energy as $E_{\mathrm{T}}=E\sin{\theta}$, and transverse momentum as $p_T=p\sin{\theta}$.
\bibitem{flavor} The term light-flavor is used for up, down, strange, and
charm flavored quarks or combinations of these.
\end{thebibliography}
\end{document}